\documentclass[aps,prl,twocolumn,superscriptaddress,showpacs,nofootinbib]{revtex4}
\usepackage{amsbsy,latexsym}
\usepackage{amsfonts}
\usepackage{amssymb}
\usepackage[mathscr]{eucal}
\usepackage{graphics}
\newcommand{\tht}{{\text{\boldmath$\theta$}}}

\newcommand{\OO}{\mathcal{O}}
\newcommand{\ord}{\mbox{\tiny$\cal O$}}

\newcommand{\aver}[1]{\langle #1 \rangle}
\newcommand{\tr}{{\rm tr}\,}

 \newcommand{\eqref}[1]{(\ref{#1})}

\begin{document}
\title{Separable Measurement Estimation of Density Matrices\\
 and its Fidelity Gap with Collective Protocols}

\author{E.~Bagan}
\affiliation{Grup de F{\'\i}sica Te{\`o}rica \& IFAE, Facultat de Ci{\`e}ncies,
Edifici Cn, Universitat Aut{\`o}noma de Barcelona, 08193 Bellaterra
(Barcelona) Spain}

\author{M. A. Ballester}
\affiliation{Centrum voor Wiskunde en Informatica, Kruislaan 413,
1098 SJ Amsterdam, The Netherlands}

\author{R.~D.~Gill}
\affiliation{Department of Mathematics, University of Leiden, Box
9512, 2300 RA Leiden, The Netherlands}
 \affiliation{EURANDOM, P.O. Box 513-5600 MB Eindhoven, The
 Netherlands}

\author{R.~Mu{\~n}oz-Tapia}
\affiliation{Grup de F{\'\i}sica Te{\`o}rica \& IFAE, Facultat de Ci{\`e}ncies,
Edifici Cn, Universitat Aut{\`o}noma de Barcelona, 08193 Bellaterra
(Barcelona) Spain}

\author{O.~Romero-Isart}
\affiliation{Grup de F{\'\i}sica Te{\`o}rica \& IFAE, Facultat de Ci{\`e}ncies,
Edifici Cn, Universitat Aut{\`o}noma de Barcelona, 08193 Bellaterra
(Barcelona) Spain}

%\date{\today}

\begin{abstract}
We show that there exists a gap between the performance of separable
and collective measurements in qubit mixed-state estimation that
persists in the large sample limit. We characterize the gap with
sharp asymptotic bounds on mean fidelity. We present an adaptive
protocol that attains the separable-measurement bound. This protocol
uses von Neumann measurements and can be easily implemented with
current technology.
\end{abstract}

\maketitle

Collective measurements provide the largest amount of information
that can be retrieved from a multipartite quantum system. However,
they rarely offer much advantage over separable (also known as {\em
unentangled}) measurements
%~\footnote{These are those where all the
%positive operators $\{O_i\}$ of the measurement can be written as a
%convex combination of local operators in the $N$ parties of the
%system, i.e. $O_i=\sum_i c_i O_{i_1}\otimes O_{i_2}\otimes \cdots
%\otimes O_{i_N}, i=i_N\cdots i_2i_1$. }
as far as state
discrimination~\cite{wootters} or estimation based on large samples
of identical copies is concerned. To be more precise, as the sample
size  $N$ goes to infinity the (mean-squared) error in the (optimal)
estimate often vanishes at the same rate for both collective and
separable measurements, despite the fact that the former are fully
general whereas the latter are much more constrained. Examples of
this abound in the literature. They include: estimation of qubit
pure states~\cite{.} (separate individual measurements suffice in
this case); estimation of spectrum of a qudit density
matrix~\cite{manuel} (for qubits see~\cite{bmo-purity});
discrimination of two non-orthogonal multipartite pure
states~\cite{virmani+plenio}; multiple-copy 2-state
discrimination~\cite{acin+us}; the problem of distinguishing the
states of the double-trine ensemble~\cite{wootters,p+w};~etc. (In
these discrimination examples $N$ needs not be asymptotically
large.)

In this letter, we meet the opposite, less common situation, but for
a very important example.  (I), we present a state-estimation
scenario where separable measurements are outperformed by collective
measurements {\em even} in the large sample limit. (II), we give a
protocol based solely on local operations and classical
communication (a so-called LOCC protocol) whose mean-squared error
(fidelity) attains the lower (upper)  separable-measurement bound.
To the best of our knowledge this is the first time a complete
example of (I) {\em and} (II) is given. The example which we provide
is of great practical relevance:  it concerns the estimation of a
qubit density matrix $\rho=(\openone+ \vec{r}\cdot \vec{\sigma})/2$,
$|\vec r|\equiv r\le1$ (the components of $\vec\sigma$ are the three
Pauli matrices) given $N$ identical copies; i.e., assuming we are
given the state $\rho^{\otimes N}$. The fact that the protocol uses
von Neumann measurements and just one-step adaptivity  adds greatly
to its practical interest, since it can be implemented with
present-day  technology. Similar protocols have been proposed
earlier in, e.g., \cite{gillmassar,Matsumoto:holbound,hayashi's
book}, but they have not been studied from the point of view of
average fidelity. Our protocol is delicately tuned to attain the
asymptotic maximum separable-measurement fidelity.

The estimation of qubit mixed states already showed some other
puzzling anomalies.  In~\cite{bbm-mixed} it was proved that the
average error using (local) tomography vanishes as~$N^{-3/4}$, which
conflicts with the behavior~$N^{-1}$ expected on statistical
grounds, as well as being much worse than the optimal collective
results~\cite{us-mixed}. A closer look at this problem reveals that
this strange power law is intimately connected to the particular but
very natural choice of prior distribution used there. This will be
explained, after introducing our notation.

Our aim is to maximize the mean fidelity $F=(1+\Delta)/2$, where
\begin{equation}\label{fidelity}
     \Delta=\sum_\chi \int dn\,dr\,w(r)\; {\bf r}\cdot{\bf R}_\chi\,
      \tr[O_\chi \rho^{\otimes N}] .
\end{equation}
In writing~(\ref{fidelity}) we have used that  (a) the fidelity
between~$\rho$ and its estimate $\rho_\chi=(\openone+\vec
R_\chi\cdot\vec\sigma)/2$ can be cast in the form
$f(\rho,\rho_\chi)=\big(\tr\sqrt{\sqrt{\rho_\chi}\rho\,
 \sqrt{\rho_\chi}}\,\big)^2=(1+{\bf r}\cdot{\bf R}_\chi)/2$,
 where the (Euclidean)
4-dimensional vector~$\bf r$ (similarly ${\bf R}_\chi$) is defined
as ${\bf r}=(\sqrt{1-r^2},\vec r\,)$; (b) a generalized measurement
on~$\rho^{\otimes N}$ is represented by a positive operator valued
measure (POVM), ${\cal O}=\{O_{\chi}\}$; (c) the prior probability
distribution of $\rho$ is isotropic ($dn$ is the rotationally
invariant measure on $\mathbb S^{d-1}$, where $d=3,2$ depending
respectively on whether  we deal with the entire Bloch sphere or
only with its equatorial plane); and (d) the purity (i.e., $r$)
prior distribution has probability density~$w(r)$.

 This letter focusses on measurements for
 which~$O_{\chi}=\bigotimes_{k=1}^{N}O^{(k)}_{\chi_k}$ (so that~$\chi
=\chi_N \chi_{N-1}\ldots \chi_1$ is a string of outcomes, each of
them associated with one of the copies of~$\rho$). By definition,
separable measurements are those whose components are convex
combinations of these ``local" operators. All LOCC measurements are
separable, but not viceversa~\cite{bennett}.

It is worth mentioning that, although we do not stick to any
particular purity prior, many arguments favor the choice of the
Bures distribution,
\begin{equation}\label{bures}
    w_{\rm Bures}(r)={2\Gamma\left({d+1\over2}\right)\over\sqrt\pi
    \Gamma\left({d\over2}\right)}{r^{d-1}\over\sqrt{1-r^2}}  .
\end{equation}
Notice that~\eqref{bures} is precisely the volume element induced by
the distance $(1/2)\arccos f(\rho,\rho+d\rho)$~\cite{bures}. It is
monotonically decreasing under coarse-graining~\cite{petz}. It has
been argued that it corresponds to maximal randomness of the signal
states and hence describes an ensemble for which one has minimal
prior knowledge~\cite{hall}.

Let us summarize some known facts about the optimal (collective)
protocols. For asymptotically large samples the maximum value of the
fidelity reads~\cite{us-mixed} (see also~\cite{Matsumoto:holbound})
\begin{equation}\label{F3D}
    F^{\rm max}_{d=3}=1-\frac{3+2 \bar{r}}{4N}+\ord(1/N),
\end{equation}
where $\bar r$ stands for the average of the purity with respect to
its prior (on which very mild regularity conditions need be
assumed), i.e., $ \bar{r} \equiv \int_0^1  dr\, w(r)\, r$.
For the Bures prior, Eq.~\eqref{bures}, we have $\bar{r}=2/(3\pi)$.
Similarly~\cite{us-mixed},
\begin{equation}\label{F2D}
F^{\rm max}_{d=2}= 1 - \frac{1}{2 N}+\ord(1/N),
\end{equation}
which is independent of $w(r)$. These asymptotic bounds were
computed assuming no restriction upon the type of measurements used
in the protocols. Although they were shown to be attained by a
specific collective measurement, one cannot rule out that separable
measurements can also attain the bounds (especially after our
introductory remarks about the frequent asymptotic optimality of
local protocols).

To get a hint for this problem, we recall some
results~\cite{bbm-mixed} concerning tomography, which is a standard
scheme for quantum state estimation in the laboratory at present. In
the simplest approach one performs  measurements of the projections
of~$\vec\sigma$ along~$d$ {\em fixed} orthogonal directions (one
does not make use of classical communication), each of them on a
corresponding fraction of the sample of size~$N/d$. For the Bures
prior the best data processing leads to
\begin{equation}\label{fidelity-tomography}
F^{\rm tom}_d=1-\frac{\xi_d}{N^{3/4}}+\ord(N^{-3/4}),\quad
0<\xi_{2}\leq\xi_{3},
\end{equation}
where the specific value of the constant $\xi_d$  is irrelevant to
this  discussion. This asymptotic behavior  contrasts drastically
with~(\ref{F3D})  and~(\ref{F2D}), and might lead us to suspect that
there exists no local estimation protocol for which~$1-F\sim N^{-1}$
if $w(r)$ is the Bures prior~(\ref{bures}). We show below that this
is not so.

But let us first provide an explanation of
the behavior shown in~(\ref{fidelity-tomography}), which can be
traced back to the contribution to $F$ from states near the surface
of the Bloch sphere (almost pure states). Let
$\mathcal{S}_\varepsilon$ be the outer shell of the Bloch sphere of
thickness~$\varepsilon$,  i.e.,
$\mathcal{S}_\varepsilon=\{\vec{r}:1-\varepsilon<r\leq 1\}$. As an
extreme case, consider the fidelity $f(\rho,\rho')$ between two
states of~${\cal S}_\varepsilon$ whose Bloch vectors point in the
same direction while one is of length~$r=1$, the other of
length~$r'=1-\varepsilon$. We find that~$1-f (\rho,\rho')\sim
\varepsilon/2$ [instead of $1-f(\rho,\rho')\sim \varepsilon^2$,
which holds for states in the complement of~${\cal S}_\varepsilon$
whose Bloch vectors  are a distance~$\varepsilon$ apart].
For a signal state not in the direction of one of the measurement
axes, however we process the tomographic data $\chi$, we cannot hope
to reconstruct the location and in particular the \emph{length}
of the Bloch vector to an accuracy better than~$N^{-1/2}$. Thus
$1-f(\rho,\rho_\chi)\sim N^{-1/2}$
when~$\rho\in\mathcal{S}_{N^{-1/2}}$. Integrating \eqref{bures} from
$r=1-\varepsilon$ to $r=1$, we see that the signal state has a
probability~$p\sim\varepsilon^{1/2}$ of being in ${\cal
S}_{\varepsilon}$. In particular, $\rho\in{\cal S}_{N^{-1/2}}$ with
probability~$p\sim N^{-1/4}$. In this case, the best processing of
the data can at most result in $1-f(\rho,\rho_\chi)\sim N^{-1/2}$.
The (dominant) contribution to $1-F$ from this increasingly thin
outer shell of signal states, ${\cal S}_{N^{-1/2}}$, is therefore of
order $[1-f(\rho,\rho_\chi)]\times p\sim N^{-3/4}$.

Let us now move on to the central part of our work and prove the
existence of a gap between the asymptotic fidelities of separable
and collective protocols. This, in turn, provides an upper bound for
the LOCC fidelity. To this end, we recall some results concerning
quantum statistical inference theory (a comprehensive summary of
general results can be found in~\cite{us-mixed}). Hereafter we write
simply~$f$ for~$f(\rho,\rho_\chi)$. From classical statistical
arguments, the average of the fidelity over the
outcomes can be expected to be
\begin{equation}\label{eq:asympmseN}
 \langle f\rangle_{\chi}=1-\frac{1}{4N}\tr\{H(\tht )
 [I(\tht, \OO)]^{-1}\}+\ord(1/N),
\end{equation}
where $\tht=(r,\theta,\phi)$ (i.e., the standard spherical
coordinates), $H(\tht)={\rm diag}[1/(1-r^2),r^2,r^2\sin^2\theta]$ is
the quantum Fisher information matrix (QFI), and $I(\tht,\OO)$ is
related to the (``classical") Fisher information matrix (FI)
$I_N(\tht,\OO)$ corresponding to a measurement~$\OO$
on~$\rho^{\otimes N}$ through the equation~$I(\tht,\OO)=\lim_{N\to
\infty}I_N(\tht,\OO)/N $. (For states on the equatorial plane,
$d=2$, we just drop the $\phi$-entry in the above expressions.) Up
to a constant factor,  $H$ is the Riemannian metric corresponding to
the fidelity  \cite{Hubner} .  The FI plays a similar role with
respect to the classical fidelity (overlap) between probability
distributions. The inverse of $H$ is a lower bound to the inverse of
the FI which is a lower bound for the variance of ``reasonable''
estimators of $\tht$ \cite{BraunsteinCaves}.

If one restricts oneself to separable measurements, the following
bound holds
\begin{equation}\label{eq:gmbound}
\tr\{[ H(\tht )]^{-1}I(\tht, \OO)\}\leq 1,
\end{equation}
as proved in~\cite{gillmassar}. It follows straightforwardly that
 \begin{equation}\label{eq:gmbound-2}
  \tr\{ H(\tht )[I(\tht, \OO)]^{-1}\}\geq d^2 .
 \end{equation}
Eqs.~\eqref{eq:gmbound-2} and (\ref{eq:asympmseN}) suggest that for
any separable measurement scheme the following bound should apply:
\begin{equation}\label{eq:f-local-bound}
\lim_{N\to\infty}N[1-\langle f\rangle_\chi]\geq \frac{d^2}{4} .
\end{equation}
One could moreover hope that the bound remains true after averaging
with respect to any prior.

A direct and rigorous proof of the desired result can be
given using exactly the same arguments as in appendix~H
of~\cite{us-mixed}. Alternatively, it follows from
a general theorem  proved
in~\cite{asqinfbd}. Either way, we have the inequality
\begin{equation}\label{eq:gap}
\lim_{N\to\infty}N (1-F^{\rm sep}_d)\ge d^2/4>\lim_{N\to\infty}N
(1-F^{\rm max}_d)
\end{equation}
where the second, strict, inequality, which  follows
from~(\ref{F3D}) and~(\ref{F2D}), proves the existence of a gap
between the two asymptotic optimal fidelities.

Within the so-called pointwise approach to quantum state estimation,
the existence of a gap between optimal collective and separable
measurements on multi-parameter problems has been known for some
time; see Hayashi and Matsumoto~\cite{Matsumoto:holbound} and their
references. In this approach, one compares the pointwise rate of
convergence (i.e., at each fixed~$\tht$), with respect to mean
square error, of estimators satisfying regularity conditions (e.g.,
``asymptotically locally unbiased"). Differing efficiencies in this
approach suggest, but {\em do not prove}, that a corresponding gap
exists when we compare \emph{average} (with respect to a prior)
\emph{fidelity}  of \emph{arbitrary} estimators.

We are now in a position to state precisely and prove our main
result:  there exists a (LOCC) one-step adaptive protocol that
saturates the separable-measurement bound~(\ref{eq:gap}). The
protocol which, taking inspiration from the Gill-Massar
approach~\cite{gillmassar}, makes use of adaptivity (and thus of
classical communication) only once, is as follows. In a first step,
we spend a vanishing fraction, $N^\alpha\equiv N_0$
($1/2<\alpha<1$), of  copies of~$\rho$ to get a rough estimate
of~$\vec n$ ($\theta$ and $\phi$), to which we refer as $\vec{n}_0$.
To this purpose we may use, e.g., tomography.

In a second step we use tomography again on the remaining
$N-N_0\equiv N_1 d$ copies of~$\rho$, but now we measure the
projection of~$\vec\sigma$ along $\vec n_0$ and along $d-1$ other
orthogonal axis in the plane normal to~$\vec n_0$. In the following
we refer to these axis as~$\vec z$, $\vec x$ and $\vec y$
respectively; they define a spatial reference frame related to the
original one through a known rotation. The outcomes of this second
step can be written as
$\chi=(\chi_x,\chi_y,\chi_z)\equiv(2\alpha_x-1,2\alpha_y-1,2\alpha_z-1)$,
where~$\alpha_x$ is the relative  frequency of plusses~($+$)
obtained in the $N_1$ measurements of $\vec x\cdot\vec\sigma$
($\alpha_y$ and $\alpha_z$ are defined similarly). The estimate of
$\vec r$ is given by $\vec R_\chi=R_\chi \vec n_\chi$, where we have
defined
\begin{eqnarray}
R_\chi&=&\chi_z,\\
\vec n_\chi&=& \vec x\sin \hat\theta \cos \hat\phi  + \vec y \sin
\hat\theta \sin \hat\phi + \vec z\cos \hat\theta
\end{eqnarray}
and
\begin{equation}
\sin\hat\theta={\sqrt{\chi_x^2+\chi_y^2}\over R_\chi},\quad
\tan\hat\phi={\chi_y\over \chi_x}.
\end{equation}
(For $d=2$ we drop the $y$-component of $\chi$ and set
$\hat\phi=0$.) This protocol is similar to the one used in
\cite{bmo-purity}, where one was only interested in estimating the
purity. The main difference  is that, in purity estimation, after
the first step one measures the rest of the copies along the
estimated direction. In the case studied here, however, part of the
copies are used to refine the estimate of the direction. The other
main difference is in the model: purity estimation is a
one-parameter model, which  essentially  behaves as a ``classical''
problem, and LOCC protocols do attain the optimal (collective)
asymptotic accuracy. In contrast, the estimation of the whole
density matrix is not classical and collective measurements can and
do provide an advantage.

Let us  prove that the fidelity of the protocol above attains the
separable bound~(\ref{eq:gap}).
 The accuracy with which $\vec n$ is estimated in step one can be quantified by the
average of $C=\cos\Theta$ over the $N_0$ outcomes of this first
measurement, where $\Theta$ is the angle between $\vec{n}$ and its
rough guess~$\vec{n}_0$. One has
\begin{equation}\label{eq:<C>_0}
\langle C \rangle_{0}=1-{\eta_d(r) \over N_0}+\ord(1/N_0),
\label{Theta}
\end{equation}
where $\eta_3(r)=3(1/r^2-1/5)$ and $\eta_2(r)=(1/r^2-1/4)$.
A~shorthand notation similar to that in~(\ref{eq:asympmseN})
and~(\ref{eq:<C>_0}) will be used below to denote other averages.
E.g.,  $\langle\cdots\rangle_{\vec r}$  will stand for the average
over the prior distribution $d\rho=dr \,w(r)dn$. Likewise, the
average fidelity (after step two) can be written as $F=\langle
f\rangle_{ \chi 0 \vec r}\equiv(1+\langle \delta \rangle_{ \chi 0
\vec r})/2$. Note that the frequencies $\alpha_{x}$, $\alpha_{y}$
and $\alpha_{z}$ are binomially distributed as
$\alpha_{x,y}\sim\text{Bin}[N_1,(1+r\,n_{x,y})/2]$
and~$\alpha_z\sim\text{Bin}[N_1,(1+r\,C)/2]$, where the components
of $\vec n$ are referred to the rotated reference frame. Hence, for
{\em large} $N_1$ ($N$), the components of $\chi$ are close to normally
distributed; $\chi_{x,y}\sim {\rm N}[r n_{x,y},N_1^{-1/2}(1-r^2
n_{x,y}^2)^{1/2}]$ and similarly~$\chi_z \sim {\rm N}[r\,
C,N_1^{-1/2}(1-r^2 C^2)^{1/2}]$.

To compute the asymptotic form of~$\Delta=\langle \delta \rangle_{
\chi 0 \vec r}$, we note that $\delta={\bf r}\cdot{\bf R_\chi}=\vec
r\cdot\vec
R_\chi+(1-r^2)^{1/2}(1-R^2_\chi)^{1/2}\equiv\delta^V+\delta^S$, as
can be read off from Eq.~(\ref{fidelity}), and make in~$\langle
\delta^V \rangle_{\chi 0 }$ (no average over the prior) the
approximation $R_\chi\approx r\, C$, along with the substitutions
$\sin\hat\theta\cos\hat\phi=\chi_x/R_\chi$ and $\sin\hat \theta \sin
\hat\phi = \chi_y /R_\chi$. Retaining only terms up to
order~$\hat\theta^2$  (on average, $\hat\theta$ is small for large
$N_1$) we have
\begin{equation}
\langle \delta^V\rangle_{\chi 0}\approx  r^2\left\langle
\!\!C^2\!-\!{\langle \chi_x^2+\chi_y^2\rangle_\chi \over2r^2}
 +{1\over r}\!\!\sum_{i=x,y} n_i\langle\chi_i\rangle_{\chi}\!\!\right\rangle_0.
\end{equation}
(For $d=2$ we just drop $\chi_y$.)

 The average
$\aver{\chi_x^2+\chi_y^2}_\chi$ (or just $\aver{\chi_x^2}_\chi$ if
$d=2$) can be  computed trivially recalling that $\chi_{x,y}$ are
almost normally distributed random variables. The resulting
expression can be written as
\begin{equation}
\langle \delta^V\rangle_{\chi 0}=
\frac{r^2}{2}\aver{(1-C)^2}_0+r^2\aver{C}_0
-\frac{d-1}{2N_1}+\ord(1/N_1) , \label{D1+D2}
\end{equation}
where we have used the relation $n_x^2+n_y^2=1-C^2$ ($n_x^2=1-C^2$
if $d=2$).

We now observe that $r^2\aver {C}_{0}$ can be approximated by~$\aver
{rR_\chi}_{\chi_z 0}$ and that the term $\aver{(1-C)^2}_0$ is of
order~$N_0^{-2}$ [see Eq.~\eqref{Theta}]. The latter is thus
subdominant if $\alpha>1/2$ and can be dropped. Therefore we obtain
(up to the order we are interested in)
\begin{eqnarray}
\Delta&=&\left\langle rR_\chi +
(1-r^2)^{1/2}(1-R_\chi^2)^{1/2}\right\rangle_{\chi_z 0 \vec r}
\nonumber
\\
&-&\frac{d-1}{2N_1}+\ord(1/N).
\end{eqnarray}
This is a gratifying result because the term in brackets can be
recognized as the average fidelity used in the purity estimation
problem discussed in \cite{bmo-purity}, and we just need to borrow
the asymptotic expression obtained there:
$F^{\text{purity}}=1-1/(2N_1)+\ord(1/N_1)$.
%The use of this
%expression is justified since this fidelity is that of an adaptive
%protocol like ours, whose second step consists of performing a
%measurement of $\vec n_0\cdot\vec\sigma$ on a number~$N_1$ of copies
%of~$\rho$.
We have
$\Delta=1-d/(2N_1)+\ord(1/N_1)=1-d^2/[2(N-N_0)]+\ord(1/N)$, and
therefore
\begin{equation}
F_d= 1-\frac{d^2}{4 N}+\ord(1/N),
\end{equation}
which is the separable-measurement bound~\eqref{eq:gap}.

Some care regarding the constant $\alpha$ that determines the
(vanishing) fraction of copies used in the first step must be taken
in the above derivation. This constant must be carefully tailored to
the specific choice of the prior~$w(r)$, Eq.~\eqref{bures}. One can
show that near $r=0$ there appears a term of order ${\cal O}(1/N^{5
\alpha/2})$ while near $r=1$ there is a term ${\cal O}(1/N^{3
\alpha/2})$ , which comes from the purity estimation
part~\cite{bmo-purity}. Therefore the choice $\alpha>2/3$ renders
both terms subdominant. The optimal value of $\alpha$ is hard to
find analytically, however numerical results suggest that it is
close to its lower bound, $2/3$.

Let us summarize. We have analyzed LOCC estimation protocols for
qubit mixed states. These are the most relevant arrangements for
practical purposes. Using statistical tools we have obtained an
asymptotic bound on the fidelity for slightly more general
approaches; those that use {\em separable} measurements. Our
specific LOCC protocol attains the separable bound. The rate at
which perfect determination can be attained is comparable to that
of the completely unrestricted optimal protocol, which involves
joint measurements: $1-F$ goes to zero at rate $N^{-1}$ for both
separable and collective approaches. The accuracies, however,
exhibit a gap, Eq.~\eqref{eq:gap}. The separable-measurement
bounds do not depend on the prior distribution. In view of the
fact that even optimal processing of standard (fixed) tomography
leads to accuracies that go to zero more slowly than $N^{-1}$ for
the very natural choice of the Bures prior, it is nontrivial and
gratifying to exhibit an experimentally feasible LOCC protocol
that saturates the separable bounds (showing they are sharp!) and
in particular has the $N^{-1}$ rate. Our results can be extended
to the distillation of pure states \cite{wip}. Also they can be
applied to higher dimensional systems, e.g., the pointwise
approach, Eqs.~(\ref{eq:asympmseN}-\ref{eq:gap}), show that the
asymptotic expression of the fidelity also satisfies $1-F=
\ord(1/N)$.

We thank A. Ac\'{\i}n, J. Calsamiglia, and W.K.~Wootters for
useful discussions. We acknowledge financial support from Spanish
Ministry of Science and Technology projects BFM2002-02588,
FIS2005-01369, CIRIT project SGR-00185, Netherlands Organization
for Scientific Research NWO project 613.003.047, the European
Community projects QUPRODIS contract no. IST-2001-38877 and RESQ
contract no IST-2001-37559. This work was done while R.D.G and
M.A.B. were at the Mathematical Institute, University of Utrecht.
%%%%%%%%%%%%%%%%%%%%%%%%%%%%%%

\end{document}